\documentstyle[12pt]{article}
\begin{document}

{\pagestyle{empty} }

\vskip 6mm

\centerline{\large \bf
Solution Independent Analysis of Black Hole Entropy }
\centerline{\large \bf
in Brick Wall Model}

\vspace{10mm}

\centerline{Masakatsu Kenmoku \footnote{E-mail address:
        kenmoku@phys.nara-wu.ac.jp} }
\centerline{\it Department of Physics,
Nara Women's University, Nara 630-8506, Japan}

\vskip 3mm
\centerline{ Kamal Kanti Nandi \footnote{E-mail address:
              kamalnandi1952@yahoo.co.in }}
\centerline{\it Department of Mathematics,
University of North Bengal}
\centerline {\it Darjeeling (W.B.) 734 430,
India }

\vskip 3mm
\centerline{Kazuyasu Shigemoto \footnote{E-mail address:
        shigemot@tezukayama-u.ac.jp} }
\centerline{\it Department of Physics,
Tezukayama University, Nara 630-8501, Japan} 

\vskip 1cm \centerline{\bf Abstract}

Using the brick wall regularization of 't Hooft, the entropy of non-extreme
and extreme black holes is investigated in a general static, spherically
symmetric spacetime. We classify the singularity in the entropy by
introducing a {\it new} index $\delta $ with respect to the brick wall
cut-off $\epsilon $. The leading contribution to entropy for non-extreme
case $(\delta \neq 0)$ is shown to satisfy the area law with quadratic
divergence with respect to the invariant cut-off $\epsilon _{{\rm inv}}$
while the extreme case $(\delta =0)$ exhibits logarithmic divergence or
constant value with respect to $\epsilon $. The general formula is applied
to Reissner-Nordstr\"{o}m, dilaton and brane-world black holes and we obtain
consistent results.

\vskip 3mm

\noindent PACS number(s): 04.70.-s, 95.30.Sf, 97.60.Lf \hfil \vfill

\newpage

%%%%%%%%%%%%%%%%%%%%%%% Section 1 %%%%%%%%%%%%%%%%%%%%%%%%

\section{Introduction}

An important part of black hole physics is concerned with its dynamical
behavior which includes studying its thermal property \cite{to:},
information content \cite{bo:}, not to mention the possible laboratory
simulation of the behavior. A basic ingredient in black hole thermodynamics
is the notion of entropy. The entropy $S$ of a standard non-extreme black
hole obeys the well known Bekenstein-Hawking area law, $S={A}/{(4G),}$ where
$A$ is the area of the horizon \cite{be:,ba:,ha:}. This result can be
derived by the action integral method for the Kerr-Newman solutions 
and de Sitter space \cite{gi:} and for dilaton black holes \cite{ka:}. The
contribution to the entropy of quantum fields in black hole backgrounds was
studied using the brick wall model \cite{'th:}, 
the WKB approximation \cite{su:} and the path integral method
\cite{al:}. However, the entropy due to
quantum fields in the black hole background introduces divergences which are
interpreted as renormalizations of the gravitational coupling constant $G$.

In the extreme case, the situation is different, that is, the area law does
not hold and the classical entropy becomes zero by the topological argument
\cite{ha-ho-ro:} and by the Gauss-Bonnet theorem \cite{tei:}. It has been
argued that extreme and non-extreme black holes should be regarded as
qualitatively different objects due to discontinuity in the Euclidean
topology. In this context, one would recall that the extreme dilaton black
holes have zero area and hence $S=0$ \cite{gi-ka:}. However, its temperature
could be zero or infinite (even arbitrary) as the Euclidean section is
smooth without identification \cite{gi-ma:}. However, $S$ does not remain
zero when quantum field contributions are taken into account: the linearly
divergent contribution vanishes but the logarithmic 
divergence persists \cite{gh:}. 
The situation is somewhat similar to the case for 
the Reissner-Nordstr\"{o}m black holes \cite{ha-ho-ro:,gh-mi:}.

Additionally, given the current widespread interest in the brane theory, it
would be important to assess similar quantum field contributions to entropy
in the background of brane-world black holes. The brane theory is described
by the Randall-Sundrum framework \cite{ra:} and the black hole solutions are
obtained by solving the Shiromizu-Maeda-Sasaki equations \cite{shi:}. The
general class of four dimensional black hole solutions has been recently
obtained by Bronnikov, Melnikov and Dehnen \cite{br:}. Typically, black
holes on the brane are characterized by induced tidal charge \cite{dmpr:},
negative energy \cite{da:}, non-singular nature \cite{sha:} and so on. In
these respects, they are different from the Reissner-Nordst\"{o}m and
dilaton black holes.

The purpose of this paper is to systematically study the quantum scalar
field contributions to the entropy using the general form of static,
spherically symmetric background spacetime. The regularization is done using
the standard 't Hooft brick wall model in evaluating the scalar field
contribution to entropy \cite{'th:}. Now, in the literature, the black hole
radiation is assumed to emerge from the inner boundary wall itself and the
Hawking temperature in the observed power spectrum followed quite
beautifully \cite{zo:}. In a similar methodological spirit, we define the
temperature at the brick wall position as the local temperature that leads
to the correct Hawking temperature in the horizon limit. With this
regularization scheme for temperature and entropy together, we want to
examine the relation between non-extreme and extreme black holes in detail.
It might be parenthetically noted that there are several other
regularization schemes too. For instance, Pauli-Villars regulator was
applied to study black hole entropy 
by Demers, Lafrance and Myers \cite{de:}. 
Also, the concept of local temperature was applied to extreme black holes
by Wang, Su and Abdalla \cite{wsa:} that yielded consistent results.
On the problem of black hole entropy, see also \cite{:others}.

The plan of the paper is as follows: Formulation of the scalar field entropy
in the brick wall model will be done in section 2. We apply the general
expression for the entropy to Reissner-Nordstr\"{o}m, dilaton and
brane-world black holes in section 3. Special interest is in the relation
between extreme, non-extreme black holes. Summary and discussion will appear
in section 4. The rationale and validity of our regularization scheme are
elucidated in the Appendix.

%%%%%%%%%%%%%%%%%%%%%%%%% Section 2 %%%%%%%%%%%%%%%%%%%%%%%

\section{Entropy of scalar field in the gravitational background}

In order to study the black hole entropy for a scalar field in the brick
wall model of 't Hooft \cite{'th:}, we start with the general diagonal
expression for the static background metric using units such 
that $\hbar=\kappa =c=G=1$ unless otherwise specified.

\subsection{The brick wall model}

General arguments imply that the metric functions depend only on radial
coordinate $r$ in (3+1) space-time dimension:
\begin{equation}
ds^{2}=g_{tt}(r){dt}^{2}+g_{rr}(r)dr^{2}+g_{\theta \theta }(r)d\Omega _{2}\ ,
\label{e1}
\end{equation}
where $d\Omega _{2}=d\theta ^{2}+\sin ^{2}{\theta }d\phi ^{2}$ is the line
element on a unit sphere. 
The matter action for the scalar field with mass $m $ is
\begin{eqnarray}
I_{M}(\phi )=-\frac{1}{2}\int d^{4}x\sqrt{-g}\left( g^{\rho \sigma }\ \partial
_{\rho}\phi \partial _{\sigma}\phi +\mu^{2}\phi^2 \right) .
\label{e2}
\end{eqnarray}
The field equation for the scalar field is obtained from this action as
\begin{eqnarray}
\frac{1}{\sqrt{-g}}\partial _{\rho}\left( \ \sqrt{-g}g^{\rho \sigma }\partial
_{\sigma}\phi \right) -\mu^{2}\phi =0.
\label{e3}
\end{eqnarray}
We may take the ansatz for the scalar field $\phi $ in the form
\begin{eqnarray}
\phi (t,r,\theta ,\varphi )=\exp ({-iEt})Y_{\ell m}(\theta ,\varphi
)f_{E,\ell }(r),
\label{e4}
\end{eqnarray}
and obtain the radial equation as
\begin{eqnarray}
\left( 
-\frac{E^{2}}{g_{tt}}
+\frac{1}{g_{\theta \theta} \sqrt{-g_{tt}g_{rr}}}
 \partial_{r} 
g_{\theta \theta} \sqrt{-\frac{g_{tt}}{g_{rr}}} 
\partial_{r} -\frac{\ell (\ell +1)}{g_{\theta \theta }}-\mu^{2}
\right)
f_{E,\ell }(r)=0.
\label{e5}
\end{eqnarray}
Here we write the radial function as
\begin{eqnarray}
f_{E,\ell }(r)\sim \exp \left( \pm i\int^{r}drk_{E,\ell }(r)\right) ,
\label{e6}
\end{eqnarray}
where the radial momentum $k_{E,\ell }$ in eq.(\ref{e6}) is obtained from
eq.(\ref{e5}) in the semi-classical approximation as
\begin{eqnarray}
k_{E,\ell }=\sqrt{g_{rr}\left( -\frac{E^{2}}{g_{tt}}
-\frac{\ell (\ell +1)}{g_{\theta \theta }}-\mu^{2}\right) }.
\label{e7}
\end{eqnarray}
The non-negative integer number of radial modes $n_{E,\ell }$ using the
semi-classical quantization condition in the brick wall model is given by
\begin{eqnarray}
\pi n_{E,\ell }=\int_{r_{h}+\epsilon }^{r_{h}+L}drk_{E,\ell }(r),
\label{e8}
\end{eqnarray}
where the brick wall distance $\epsilon $ denotes an ultraviolet cutoff from
$r_{h}$ and radius $L$ denotes an infrared cutoff measured from $r_{h}$ .
The total number of states with energy $E$ is given by
\begin{eqnarray}
g(E)=\sum_{\ell ,\ell _{z}}n_{E,\ell }\simeq \int_{0}^{\ell _{max}}d\ell
(2\ell +1)\frac{1}{\pi }\int_{r_{h}+\epsilon }^{r_{h}+L}drk_{E,\ell }(r).
\label{e9}
\end{eqnarray}

To study the thermodynamics of black holes, we make the Wick rotation from
Minkowski time $t$ to Euclidian time $\tau =it$ so that 
the metric eq.({\ref{e1}}) becomes
\begin{eqnarray}
ds^{2}=g_{\tau \tau }(r){d\tau }^{2}+g_{rr}(r)dr^{2}
+g_{\theta \theta }(r){d\Omega }_{2},
\label{e10}
\end{eqnarray}
where $g_{\tau \tau }=-g_{tt}$ is the Euclidean time component of the
metric. The free energy $F(\beta )$ at the inverse temperature $\beta =1/T$
is obtained by the partition function $Z_{M}(\beta )$ for the matter action
in eq.(\ref{e2}) as
\begin{eqnarray}
Z_{M}(\beta ) &=&\int_{\beta }[d\phi ]{\rm e}^{-I_{M}(\phi )}\ ,  \nonumber
\\
F(\beta ) &=&-\frac{1}{\beta }\ln Z_{M}(\beta )
=-\int_{0}^{\infty }dE\frac{g(E)}{{\rm e}^{\beta E}-1}.  
\label{e11}
\end{eqnarray}
The integration in eq.(\ref{e11}) with respect to angular momentum $\ell $
can be performed using eq.(\ref{e9}) and the free energy is expressed in
the form
\begin{eqnarray}
F(\beta )=-\frac{1}{\pi }\int_{0}^{\infty }dE
\frac{1}{{\rm e}^{\beta E}-1}\int_{r_{h}
+\epsilon }^{r_{h}+L}dr\frac{2}{3}(g_{rr})^{1/2}g_{\theta \theta
}\left( \frac{E^{2}}{g_{\tau \tau }}-\mu^{2}\right) ^{3/2}.
\label{e12}
\end{eqnarray}
If the mass of the scalar field is zero, then the expression of the free
energy becomes simpler, viz.,
\begin{eqnarray}
F(\beta )=-\frac{2}{45\pi }
\left( \frac{\pi }{\beta }\right)^{4}\frac{V}{4\pi },
\label{e13}
\end{eqnarray}
where $V$ denotes the volume of optical space defined as
\begin{eqnarray}
V:=4\pi \int_{r_{h}+\epsilon }^{r_{h}+L}dr\left( g_{\tau \tau }\right)^{-3/2}
\left( g_{rr}\right)^{1/2}g_{\theta \theta }.
\label{e14}
\end{eqnarray}
Then the entropy of the black holes is obtained in a compact form
\begin{eqnarray}
S:=\beta ^{2}\frac{\partial F(\beta )}{\partial \beta }
=\frac{1}{45}\left(
\frac{2\pi }{\beta }\right) ^{3}\frac{V}{4\pi }.
\label{e15}
\end{eqnarray}
This result coincides with that obtained by the path integral method 
by de Alwis and Ohta \cite{al:}. We need to evaluate the temperature on the
horizon in order to obtain the value of entropy in eq.(\ref{e15}). The
temperature is defined by the condition that no conical singularity is
imposed on the Euclidean Rindler space. While this prescription consistently
yields a unique temperature of the horizon for non-extreme black holes, it
fails for the extreme case. Nevertheless, we introduce the ultraviolet
regularization $\epsilon $ and uniformly define the temperature that should
be valid for extreme as well as the non-extreme case, that is, we define the
local temperature \cite{wsa:}:
\begin{eqnarray}
\left. \frac{2\pi }{\beta }
=\frac{\partial _{r}g_{\tau \tau }}
{2\sqrt{g_{\tau \tau }g_{rr}}}\right\vert _{r=r_{h}+\epsilon }.
\label{e16}
\end{eqnarray}
Note that the regularization position is not just anywhere off the horizon
but at the boundary position of the brick wall infinitesimally close to the
horizon. The background (Hawking) temperature and gravitational entropy of
the horizon (at $r=r_{h}$) are consistently determined from the above
equation (with $\epsilon \rightarrow 0$) both for non-extreme and extreme
black holes. The meaning of this temperature regularization is further
illustrated in the Appendix in the extreme Reissner-Nordstr\"{o}m case.

%%%%%%%%%%%%%%%%%%%%%%%%%%%%%%%%%%%%%%%%%%%%%%%%%%%%%%%%%%%%%%%%%%

\subsection{Laurent expansion around the horizon}

In order to evaluate the entropy formula of eq.(\ref{e15}) independently of
any explicit form of black hole solutions, we make Laurent expansion of the
metric around the horizon $r\sim r_{h}$ as
\begin{eqnarray}
g_{\tau \tau }(r) &=&(r-r_{h})^{a}\sum_{i=0}^{\infty
}a^{(i)}(r_{h})(r-r_{h})^{i}\ ,  \nonumber \\
g_{rr}(r) &=&(r-r_{h})^{b}\sum_{i=0}^{\infty }b^{(i)}(r_{h})(r-r_{h})^{i}\ ,
\label{e17} \\
g_{\theta \theta }(r) &=&(r-r_{h})^{c}\sum_{i=0}^{\infty
}c^{(i)}(r_{h})(r-r_{h})^{i}\ ,  \nonumber
\end{eqnarray}
where the coefficients $a^{(0)},b^{(0)},c^{(0)}$, respectively, are assumed
not to be zero and $a,b,c$ denote the leading exponents of Laurent expansion
for each metric function. In the following, we keep only the ultraviolet
singularity on the horizon, which is the main contribution to entropy.
%%%%%%%%%%%%%%%%%%%%%%%%%%%%%%%%%%%%%%%%%%%%%%%%%%%%%

\begin{itemize}
\item[A.]

{\large Leading contribution}\newline
The volume of optical space in eq.(\ref{e14}) is evaluated using the
expansion in eq.(\ref{e17}), and the leading term of the contribution is
obtained as
\begin{eqnarray}
\frac{V^{(0)}}{4\pi }=\left\{
\begin{array}{ll}
(a^{(0)})^{-3/2}(b^{(0)})^{1/2}c^{(0)}\log {(L/\epsilon )} &
\mbox{if
$\gamma= 0$} \\
(a^{(0)})^{-3/2}(b^{(0)})^{1/2}c^{(0)}\ {\gamma }^{-1}\epsilon ^{-\gamma } & 
\mbox{otherwise }\ ,
\end{array}
\right.
\label{e18}
\end{eqnarray}
where we introduce $\gamma $ as the index of exponent of the optical volume
\begin{eqnarray}
\gamma :=\frac{3a}{2}-\frac{b}{2}-c-1\ .
\label{e19}
\end{eqnarray}
The leading contribution to the local temperature $T=1/\beta $ at the brick
wall distance $\epsilon $ is obtained from eq.(\ref{e16}) as
\begin{eqnarray}
\frac{2\pi }{\beta ^{(0)}}=\frac{a}{2}(a^{(0)})^{1/2}(b^{(0)})^{-1/2}
\epsilon ^{a/2-b/2-1}\ .
\label{e20}
\end{eqnarray}
The leading contribution to the entropy in eq.(\ref{e15}) is obtained by
combining the expressions in eqs.(\ref{e18}) and (\ref{e20})
\begin{eqnarray}
S^{(0)}=\left\{
\begin{array}{ll}
\displaystyle{\ \frac{1}{45}\left( \frac{a}{2}\right)
^{3}(b^{(0)})^{-1}c^{(0)}\log {(L/\epsilon )}\ \epsilon ^{-\delta }} &
\mbox{if
$\gamma=0$} \\
\displaystyle{\ \frac{1}{45}\left( \frac{a}{2}\right)
^{3}(b^{(0)})^{-1}c^{(0)}\ \gamma ^{-1}\epsilon ^{-\delta }} & 
\mbox{otherwise}\ ,
\end{array}
\right.
\label{e21}
\end{eqnarray}
where we introduce $\delta $ as a new index of exponent of entropy
\begin{eqnarray}
\delta :=b-c+2\ .
\label{e22}
\end{eqnarray}

%%%%%%%%%%%%%%%%%%%%%%%%%%%%%%%%%%%%%%%%%%%%%%%%

\item[B.] {\large Area law}\newline
By defining the area of the brick wall surface
\begin{eqnarray}
A=\int d\Omega_{2}\,g_{\theta \theta }\left. 
{}\right\vert_{r=r_{h}+\epsilon }\ ,
\label{e23}
\end{eqnarray}
and the invariant distance to the brick wall
\begin{eqnarray}
\epsilon _{{\rm inv}}:=\int_{r_{h}}^{r_{h}+\epsilon }dr(g_{rr})^{1/2}\ ,
\label{e24}
\end{eqnarray}
which holds for $b+2>0$, we obtain the area law for the leading contribution
to the entropy from eq.({\ref{e21}})
\begin{eqnarray}
S^{(0)}=\frac{1}{45}\left( \frac{a}{2}\right) ^{3}\left( \frac{2}{b+2}
\right) ^{2}\,\gamma ^{-1}(\epsilon _{{\rm inv}})^{-2}\,\frac{A}{4\pi }\ 
\mbox{ if $0<\delta$ and $\gamma\neq 0$}\ .
\label{e25}
\end{eqnarray}
The leading contribution to entropy for $\delta >0$ is proportional to the
area but diverge in $\epsilon $ (quadratically divergent 
in $\epsilon _{{\rm inv}}$). 
This term is considered as the renormalization effect to the
gravitational constant $G$ in the Bekenstein-Hawking 
entropy $\displaystyle{S_{BH}={A}/{(4G),}}$ and is the same as 
the one suggested by Susskind and Uglum \cite{su:}. 
Note that the leading contribution to entropy for $\delta<0$ is zero 
in the limit $\epsilon \rightarrow 0$.

%%%%%%%%%%%%%%%%%%%%%%%%%%%%%%%%%%%%%%%%%%%%%%%%%%%%%
\item[C.] {\large Extreme case}\newline
We define the extreme case as $\delta =0$ in which case the leading
contribution to the entropy doesn't show the area law. This contribution can
directly be read off from eq.({\ref{e21}})
\begin{eqnarray}
S^{(0)}=\left\{
\begin{array}{ll}
\displaystyle{\ \frac{1}{45}\left( \frac{a}{2}\right)
^{3}(b^{(0)})^{-1}c^{(0)}\log {(L/\epsilon )}} &
\mbox{if $\delta=0$ and
$\gamma=0$} \\
\displaystyle{\ \frac{1}{45}\left( \frac{a}{2}\right)
^{3}(b^{(0)})^{-1}c^{(0)}\ \gamma ^{-1}} &
\mbox{if
$\delta=0$ and $\gamma\neq 0$
}.
\end{array}
\right.
\label{e26}
\end{eqnarray}
The equation ({\ref{e26}}) shows that the leading contribution to entropy
for $\delta =0$ is constant or logarithmically divergent in $\epsilon $ and
is not proportional to the area on the horizon. 
The coefficient term $(b^{(0)})^{-1}c^{(0)}$ is shown to be 
a numerical factor from the
dimensional consideration. This term is considered as the renormalization of
the quadratic-curvature coupling constant in the one-loop effective
gravitational action suggested by Demers, Lafrance and Myers \cite{de:}.

%%%%%%%%%%%%%%%%%%%%%%%%%%%%%%%%%%%%%%%%%%%%%%%%%%%%

\item[D.] {\large Next leading contributions}\newline
The next leading contribution to the black hole entropy $S^{(1)}$ is from
the optical volume $V^{(1)}$ and the temperature $1/\beta ^{(1)}$. They are
calculated to be
\begin{eqnarray}
\frac{V^{(1)}}{4\pi }=\left\{
\begin{array}{ll}
\displaystyle{\ (a^{(0)})^{-3/2}(b^{(0)})^{1/2}c^{(0)}\ X\ 
\log {(L/\epsilon )}} & \mbox{if $\gamma= 1$} \\
\displaystyle{\ (a^{(0)})^{-3/2}(b^{(0)})^{1/2}c^{(0)}\ X\ 
\frac{\epsilon^{-\gamma +1}}{\gamma -1}} & \mbox{if $\gamma\neq 1$}\ ,
\end{array}
\right.
\label{e27}
\end{eqnarray}
with
\begin{eqnarray}
X:=-\frac{3a^{(1)}}{2a^{(0)}}+\frac{b^{(1)}}{2b^{(0)}}
+\frac{c^{(1)}}{c^{(0)}}\ .
\label{e28}
\end{eqnarray}
Similarly,
\begin{eqnarray}
\frac{2\pi }{\beta ^{(1)}}=\frac{a}{2}(a^{(0)})^{1/2}(b^{(0)})^{-1/2}\ Y\
\epsilon ^{a/2-b/2}\ ,
\label{e29}
\end{eqnarray}
with
\begin{eqnarray}
Y:=(\frac{1}{2}+\frac{1}{a})\frac{{a}^{(1)}}{a^{(0)}}
-\frac{1}{2}\frac{{b}^{(1)}}{b^{(0)}}\ .
\label{e30}
\end{eqnarray}
Combining them we obtain the next leading contribution to the entropy as
\begin{eqnarray}
S^{(1)} &=&\frac{2\pi ^{2}}{45}\left( (\beta ^{(0)})^{-3}V^{(1)}
+3(\beta^{(0)})^{-2}(\beta ^{(1)})^{-1}V^{(0)}\right)   \nonumber \\
&=&\frac{1}{45}(\frac{a}{2})^{3}\ {b^{(0)}}^{-1}c^{(0)}
\epsilon^{-\delta +1}Z\ ,  \label{e31}
\end{eqnarray}
where the function $Z$, given below, shows logarithmic divergence or
constant value depending on the value of index $\gamma $:
\begin{eqnarray}
Z:=\left\{
\begin{array}{ll}
\displaystyle{-X+3Y\log {\frac{L}{\epsilon }}} & \mbox{if $\gamma=0$} \\
\displaystyle{\ X\log {\frac{L}{\epsilon }}+3Y} & \mbox{if $\gamma=1$} \\
\displaystyle{\ X\frac{1}{\gamma -1}+3Y\frac{1}{\gamma }} &
\mbox{if
$\gamma\neq 0,1 $}\ .
\end{array}
\right.
\label{e32}
\end{eqnarray}
We can immediately derive the conclusion that the next leading contribution
to entropy $S^{(1)}$ is zero if $\delta <1$, logarithmically divergent or
constant if $\delta =1$ and shows other type of divergence if $\ \delta >1$,
which are derived from eq.(\ref{e31}).
\end{itemize}

%%%%%%%%%%%%%%%%%%%%%%%%%%%%%%%%%%%%%%%%%%%%%%%%%%%%

\subsection{Classification of scalar field entropy in terms of indices}

Combining the leading and next leading contributions from the ultraviolet
region of the horizon, we can classify the scalar field entropy in the black
hole background from eqs.(\ref{e21}) and (\ref{e31}) according to the two
indices $\gamma $ in eq.(\ref{e19}) and $\delta $ in eq.(\ref{e22}). This
is summarized in Table 1. \vspace{3mm}\newline
{\bf Table 1.} Classification of the black hole entropy in terms of the
indices \vspace{3mm} \newline
\noindent
\begin{tabular}{|l||l|l|l|l|}
\hline
\ \ \ \ \ Type & \multicolumn{2}{c|}{Index} & \multicolumn{2}{c|}{Entropy}
\\ \cline{2-5}
& $\delta $ & $\gamma $ & $S^{(0)}$ & $S^{(1)}$ \\ \hline\hline
(i) Regular & $\delta <0$ &  & $0$ & $0$ \\ \hline
(ii) Extreme & $\delta =0$ & $\gamma =0$ & $\log (L/\epsilon )$ & $0$ \\
&  & $\gamma \neq 0$ & constant & $0$ \\ \hline
(iii) Intermediate & $0<\delta <1$ & $\gamma =0$ & $\log (L/\epsilon
)\epsilon ^{-\delta }$ & $0$ \\
&  & $\gamma \neq 0$ & $\epsilon ^{-\delta }$ & $0$ \\ \hline
(iv) Schwarzschild- & $\delta =1$ & $\gamma =1$ & linear in $1/\epsilon $ & $%
\log (L/\epsilon )$ \\
type &  &  &  & and/or constant \\
&  & $\gamma \neq 1$ & linear in $1/\epsilon $ & constant \\ \hline
(v) More singular & $\delta >1$ &  & more singular & more singular \\
&  &  & than linear & than logarithm \\ \hline
\end{tabular}

\vspace{3mm} \noindent We remark that the leading contribution to the
entropy $S^{(0)}$ in Table 1 satisfies the area law if $\ \delta >0$ and $%
\gamma \neq 0$ as shown in eq.(\ref{e25}).

%%%%%%%%%%%%%%%%%%%%%%%%%%%%%%%%%%%%%%%%%%%%%

\section{Examples}

In this section we apply our model independent analysis of the black hole
entropy for a scalar field in the brick wall model to some special black
hole solutions. Examples that we consider are standard Schwarzschild,
Reissner-Nordstr\"{o}m, dilaton black hole and a typical example of
brane-world black hole solutions, the corresponding metrics being listed in
Table 2. \vspace{3mm}\newline
{\bf Table 2.} Metrics of black hole solutions \vspace{3mm}\newline
\noindent
\begin{tabular}{|l||l|l|l|}
\hline
\ \ Black hole solution & \multicolumn{3}{c|}{Metric} \\ \cline{2-4}
& $g_{\tau \tau }$ & $g_{rr}$ & $g_{\theta \theta }$ \\ \hline\hline
Schwarzschild & $\displaystyle{1-{2M}/{r}}$ & $\displaystyle{\frac{1}{1-2M/r}%
}$ & $r^{2}$ \\ \hline
Reissner-Nordstr\"{o}m & $\displaystyle{1-{2M}/{r}+Q^{2}/r^{2}}$ & $%
\displaystyle{\frac{1}{1-2M/r+Q^{2}/r^{2}}}$ & $r^{2}$ \\ \hline
dilaton  & $\displaystyle{1-{2M}/{r}}$ & $\displaystyle{\frac{1}{1-2M/r}}$ &
$r(r-p_{0})$ \\ \hline
brane world  & $\displaystyle{1-{2M}/{r}}$ & $\displaystyle{\frac{1-3M/2r}{%
(1-2M/r)(1-q_{0}/r)}}$ & $r^{2}$ \\ \hline
\end{tabular}

\vspace{3mm} \noindent In Table 2, $M$ and $Q$ denote the black hole mass
and charge respectively, and $p_{0}$ and $q_{0}$ are parameters. We discuss
these examples in the non-extreme and extreme cases separately.

%%%%%%%%%%%%%%%%%%%%%%%%%%%%%%%%%%%%%%%%%%%%%%%%%%%%

\subsection{Non-extreme black holes}

Non-extreme black hole is defined as $M\neq Q$ for
Reissner-Nordstr\"{o}m, $p_{0}\neq 2M$ for dilaton 
and $q_{0}\neq 2M$ for the brane-world black holes. In
all cases, the leading exponents of metrics defined in eq.(\ref{e17}) are
of the same value $a=1,b=-1$ and $c=0$. This result leads the indices
defined in eqs.(\ref{e19}) and (\ref{e22}) to have values $\gamma =1$ 
and $\delta =1$, which case we called the Schwarzschild-type in Table 1. The
reason is that, for all non-extreme black hole solution of this type, the
entropy behaves like that for the Schwarzschild solution with respect to the
ultra violet cut-off $\epsilon $ although the coefficient functions depend
on the example chosen, but the area law holds.

The explicit corrections to entropy in the Schwarzschild-like black hole
backgrounds are obtained by using the coefficient functions from the general
Laurent expansion in eq.(\ref{e17}) for each black hole solution listed in
Table 2.. The leading and next leading contributions to entropy in each
non-extreme background solution are given in the following.

\begin{itemize}
\item[E1.] {\large Non-extreme Reissner-Nordstr\"{o}m black hole}
\begin{eqnarray}
S^{(0){\rm RN}} &=&\frac{r_{+}-r_{-}}{360\ \epsilon }\ ,  \nonumber \\
S^{(1){\rm RN}} &=&\frac{1}{360\ r_{+}}
\left\{ 2(2r_{+}-3r_{-})\log \frac{L}{\epsilon }-6(r_{+}-2r_{-})\right\} 
\ ,  \label{e33}
\end{eqnarray}
where $r_{\pm }=M\pm \sqrt{M^{2}-Q^{2}}$ are the outer and inner horizons.

\item[E2.] {\large Non-extreme dilaton black hole}
\begin{eqnarray}
S^{(0){\rm dilaton}} &=&\ \frac{r_{h}-p_{0}}{360\ \epsilon }\ ,  \nonumber \\
S^{(1){\rm dilaton}} &=&\frac{1}{360\ r_{h}}\left\{ (4r_{h}-3p_{0})\log
\frac{L}{\epsilon }-6(r_{h}-p_{0})\right\} \ ,  \label{e34}
\end{eqnarray}
where $r_{h}=2M$ and $p_{0}$ are the horizon and parameter of dilaton black
hole respectively.

\item[E3.] {\large Non-extreme brane-world black hole}
\begin{eqnarray}
S^{(0){\rm BWBH}} &=&\ \frac{r_{h}-q_{0}}{90\ \epsilon }\ ,  \nonumber \\
S^{(1){\rm BWBH}} &=&\frac{1}{180\ r_{h}}
\left\{ (11r_{h}-12q_{0})\log \frac{L}{\epsilon }-3(7r_{h}-8q_{0})\right\} 
\ ,  \label{e35}
\end{eqnarray}
where $r_{h}=2M$ is the horizon and $q_{0}$ is the parameter of brane world
black hole respectively.
\end{itemize}

Clearly, the linear and logarithmic divergences in each of the above cases
are similar. Each Schwarzschild-like black hole tends to Schwarzschild black
hole if $Q=0$ for Reissner-Nordstr\"{o}m, $p_{0}=0$ for dilaton 
and $q_{0}=3M/2$ for the brane-world black hole respectively.

%%%%%%%%%%%%%%%%%%%%%%%%%%%%%%%%%%%%%%%%%%%%%%%%%%%%

\subsection{Extreme black holes}

Next we consider the extreme cases, which are $M=Q$ 
for Reissner-Nordstr\"{o}m, $p_{0}=2M$ for dilaton 
and $q_{0}=2M$ for brane-world black holes. For
each solution the corresponding leading exponent of the Laurent expansion
and the indices deriving from them are listed in the Table 3. 
\vspace{3mm}
\newline
{\bf Table 3.} Leading exponents in eq.(\ref{e17}) 
and index of the optical volume in eq.(\ref{e19}) 
and that of the entropy in eq.( \ref{e22}) 
for the extreme black holes 
\vspace{3mm}\newline
\noindent
\begin{tabular}{|ll||c|c|c|c|c|}
\hline
\multicolumn{2}{|c||}{Extreme black hole} & \multicolumn{3}{c|}{Exponent} &
\multicolumn{2}{c|}{Index} \\ \cline{3-7}
\multicolumn{2}{|c||}{} & $\ \ a\ \ $ & $\ \ b\ \ $ & $\ \ c\ \ $ & $\ \
\gamma \ \ $ & $\ \ \delta \ \ $ \\ \hline\hline
\ Reissner-Nordstr\"{o}m & $\ M=Q\ $ & 2 & -2 & 0 & 3 & 0 \\ \hline
\ dilaton & $p_{0}=2M$ & 1 & -1 & 1 & 0 & 0 \\ \hline
\ brane-world & $q_{0}=2M$ & 1 & -2 & 0 & 3/2 & 0 \\ \hline
\end{tabular}
\vspace{3mm}\newline
For all extreme cases, the values of the index $\delta $ in Table 3 show
zero. The corresponding leading contributions to entropy are either
logarithmically divergent or constant depending on the value of another
index $\gamma $, which can be read off from eq.(\ref{e26}).

The explicit expressions obtained from the metric form in Table 2. are
\begin{eqnarray}
S^{(0){\rm Ext-RN}}=S^{(0){\rm Ext-BWBH}}=\frac{1}{135}
\label{e36}
\end{eqnarray}
for the extreme Reissner-Nordst\"{o}m and brane-world black holes, and
\begin{eqnarray}
S^{(0){\rm Ext-dilaton}}=\frac{1}{360}\log {\frac{L}{\epsilon }}
\label{e37}
\end{eqnarray}
for the extreme dilaton black hole. The optical volumes 
for Reissner-Nordstr\"{o}m and brane-world black holes diverge 
like $\epsilon ^{-3}$ and $\epsilon ^{-3/2}$ 
but their temperatures tend to zero as $\epsilon $ 
and $\epsilon ^{1/2}$ respectively, and combining them the entropies become
finite as in eq.(\ref{e36}) . For dilaton black hole, the optical volume
diverges logarithmically so that the entropy also becomes logarithmically
divergent as in eq.(\ref{e37}). Note that the next leading term to entropy
from the ultraviolet contribution vanishes for the extreme black holes, that
is, $S^{(1)}=0.$

Earlier calculations 
for the extreme Reissner-Nordstr\"{o}m case \cite{gh-mi:} have
shown that $S^{(0){\rm Ext-RN}}$ 
$[=8\pi^3M^6/(135\beta^3 \epsilon^3)]$
is either zero if 
$\beta \rightarrow \infty $
or cubically divergent in the cut-off (as $\epsilon ^{-3}$) 
if the period $\beta $ is left arbitrary. 
Clearly, there is no unique answer here.
Therefore, one good way to dismiss the ambiguity due to $\beta $ is 
to regularize both the temperature and the optical volume wih 
the same regularization parameter $\epsilon$.
The brick wall regularization gives $\beta \propto \epsilon ^{-1}$ 
which is what 
we have been actually doing (see the Appendix for details). 
The constant value thereby obtained for $S^{(0){\rm Ext-RN}}$
in eq.(\ref{e36}) is perfectly consistent with the cited expression 
from \cite{gh-mi:}
in the square bracket above under necessary dimensional readjustments.
Similar considerations apply for $S^{(0){\rm Ext-BWBH}}$. 

%%%%%%%%%%%%%%%%%%%%%%%%%%%%%%%%%%%%%%%%%%%%%%%%%%%%

\subsection{Extreme black holes vs extremal limit of non-extreme black holes}

As mentioned earlier, extremal black holes (EBH) are topologically
dissimilar to their non-extreme counterparts. It is not possible to obtain
properties of EBH by continuously extremalizing their non-extreme partner.
These are the well known Hawking-Horowitz-Ross EBH \cite{ha-ho-ro:} for 
which the gravitational entropy is zero. See also \cite{tei:}. 
(There is a second kind of EBH
due to Zaslavskii \cite{za:} in which the topology is assumed 
to belong to the non-extreme sector and as a result the entropy satisfies the
Bekenstein-Hawking area law. A consistent grand canonical ensemble approach
\cite{bbwy:} however does not lead to the area law \cite{gm:}.)

We have been concerned here with the contribution to gravitational entropy
due to a scalar field in a fixed gravitational background. We state here how
these corrections in the background of EBH contrast with those in the
extreme limit of the non-extreme black hole background. The result is
solution dependent. The limits of entropy for the non-extreme 
Reissner-Nordstr\"{o}m and brane-world black holes in 
eqs.(\ref{e33}) and (\ref{e35}) are
logarithmically divergent and give different results from those for the
extreme black holes in eq.(\ref{e36}). The limit of entropy for the
non-extreme dilaton black hole in eq.(\ref{e34}) gives the same value as in
the extreme dilaton black hole in eq.(\ref{e37}). The reason is that
additional poles appear in Laurent expansion in eq.(\ref{e17}) for RN and
BWBH but gives additional zero for the dilaton black hole in the limiting
procedure.

The leading contributions to the entropy which are either logarithmically
divergent or constant can be absorbed in the renormalization of the
quadratic-curvature coupling constant in the one-loop effective
gravitational action \cite{de:}.

\section{Summary and discussion}

The merits of the present paper lie in the (a) general expression for
entropy of the scalar field, eq.(\ref{e25}), applicable to a wide variety of
spherically symmetric fixed backgrounds, (b) introduction of a new parameter
$\delta $, eq.(\ref{e22}), leading to the classifications and (c) in 
the easiness in which the higher order corrections, 
eqs.(\ref{e31}), (\ref{e32}), can be computed. 
We have also exemplified the validity of our method with some known solutions,
the brane-world case being a new addition. The fact that $S^{(1)}=0$ in the
regular, intermediate and extreme cases (Table 1) is an interesting result.

Entropy is composed of the optical volume $V$ and the (local)
temperature $1/{\beta }$, 
both of which are evaluated on the surface at the brick wall
distance $\epsilon $ from the horizon $r_{h}$. Physically, the local
temperature here is analogous to the one originally devised by Pretorius,
Vollick and Israel \cite{pvi:} and later used by others, 
for instance, \cite{wsa:}.
There the temperature is defined on a reversible contracting thin shell
gradually zeroing on the horizon while being in thermal equilibrium with
acceleration radiation. Mathematically, the local temperature on the wall
surface approaching the Hawking temperature on the horizon can be likened to
test functions approaching a generalized function (Gaussian distributions
tending to Dirac delta function, for instance). This underlines the spirit
in the definition of a local temperature as well as in the overall approach.

This kind of approach has allowed us to classify the singularity in entropy
by an index ${\delta }$ with respect to $\epsilon$. The leading 
contribution to entropy for non-extreme case $(\delta \neq 0)$ is shown to
satisfy the area law with quadratic divergence with respect to the invariant
cut-off $\epsilon _{inv}$. For the extreme case $(\delta =0)$, the
contribution is either logarithmically divergent or constant with respect to
$\epsilon $. The results are displayed in Tables 1 and 3.

The general result is applied to Reissner-Nordstr\"{o}m, brane-world and
dilatonic black holes. We have compared the extreme black holes with the
extremal limit of non-extreme black holes. The limits of entropy for the
non-extreme Reissner-Nordstr\"{o}m and brane-world black holes show
logarithmic divergence while the corresponding extreme cases produce a
constant value. These results are consistent given the fact that the
Euclidean topologies in the two situations are different. Note that the
behavior of the two solutions, viz., RN and brane-world black hole, in this
respect is very similar. This is probably due to the fact that both the
solutions satisfy the same field equation $R=0$ where $R$ is the Ricci
scalar.

On the other hand, the limit of entropy for the non-extreme dilaton black
hole gives the same result as for the extreme dilaton black hole, {\it both}
being logarithmically divergent (see \cite{gh:}) unlike in the other
solutions. Does it imply that the background dilatonic spacetime is
intrinsically different from those of other black holes? We do not know, but
would point to a certain difference: In the extreme dilatonic case, the
background gravitational entropy is zero due to the fact that the area is
zero. In the other two extreme cases, the background entropy is also zero
but the area is nonzero. It would be of interest to examine this issue in
more detail and we leave it for future investigations.

\vspace{5mm}
%%%%%%%%%%%%%%%%%%%%%%%% Acknowledgments %%%%%%%%%%%%%%%%
\begin{center}
{\Large \bf Acknowledgments}
\end{center}

It is a great pleasure to thank Parthasarathi Mitra for critical
comments 
and Guzel Kutdusova for pointing out some useful references.

\vspace{5mm}
%%%%%%%%%%%%%%%%%%%%%%%% Appendix %%%%%%%%%%%%%%%%%%%%%%%%%
\begin{center}
{\Large \bf Appendix}
\end{center}
\appendix \setcounter{equation}{0} 
\renewcommand{\theequation}{A.\arabic{equation}}
%%%%%%%%%%%%%%%%%%%%%%%%%%%%%%%%%%%%%%%%%%%%%%%%%%%%%%%%%%
\section{Regularization method for the temperature}

Why do we need to consider this temperature regularization at all? The
reason that has prompted us is this: For EBH, Hawking temperature is
completely {\it arbitrary} because the period can be identified with any
value. It is assumed that the black hole entropy $S=M{dS}/{dM}-I$ can
not depend on arbitrary temperature, and the only conclusion is to set the
classical action $I$ $\propto 1/T$ to zero \cite{tei:}. This leads to $S=kM$, 
where $k$ is an arbitrary constant. There is indeed no clear cut reason 
as yet to set $k$ to zero \cite{gh-mi:}, but if we still do it, we 
get $S=0$ for the background
EBH entropy. With this scenario in view, and fixing $S$ to its classical
zero value, we attempted to try out an alternative method for semiclassical
corrections to entropy involving a well defined temperature.

Regularization method for the temperature involves defining the (local)
temperature at a surface, infinitely close to the horizon, where it has a
definite value and after the calculation of entropy, the limit is taken to
the original place, the horizon. Now, the brick wall boundary provides a
natural surface close to the horizon. The temperature so defined is the same
as the temperature defined by the condition that no conical singularity
appears in the Euclidean Rindler space. We illustrate this by the extreme RN
case in detail.

The Euclideanized extreme RN metric is
\begin{eqnarray}
ds^{2}=(1-M/r)^{2}d\tau ^{2}+\frac{1}{(1-M/r)^{2}}dr^{2}
+{\mbox{\rm (angle term)}}\ .
\label{a1}
\end{eqnarray}
Let us define a new radial variable $R$ as
\begin{eqnarray}
R:=\int_{r_{h}}^{r}dr\frac{1}{{1-M/r}}\ .
\label{a2}
\end{eqnarray}
However the integration diverges at the horizon $r_{h}=M$. 
To avoid this, the regularization is taken at the boundary 
position of quantum scalar field $r_{h}+\epsilon $ as
\begin{eqnarray}
R:=\int_{r_{h}+\epsilon }^{r+\epsilon }dr\frac{1}{{1-M/r}}\simeq
\frac{M}{r_h+\epsilon-M} \int_{r_{h}+\epsilon }^{r+\epsilon }dr 
\simeq \frac{M}{\epsilon}(r-M).
\label{a3}
\end{eqnarray}
Note that $R$ tends to zero as $r\rightarrow r_{h}$. The original radial
coordinate $r$ is expressed inversely as
\begin{eqnarray}
r-M\simeq \epsilon \frac{R}{M}.
\label{a4}
\end{eqnarray}
Then the Euclidean component of metric is expressed by the new variable
\begin{eqnarray}
g_{\tau \tau }=(1-M/r)^{2}\simeq \frac{(r-M)^{2}}{M^{2}}
\simeq \frac{\epsilon ^{2}R^{2}}{M^{4}}\ .
\label{a5}
\end{eqnarray}
The Rindler form of the metric is therefore
\begin{eqnarray}
ds^{2}={R^{2}}(\frac{\epsilon d\tau }{M^{2}})^{2}+dR^{2}
+{\mbox{\rm (angle term)}}\ .
\label{a6}
\end{eqnarray}
No conical singularity condition is imposed to this Rindler metric 
when the Euclidean time varies [$0,\beta =1/T$], and obtain
\begin{eqnarray}
\frac{2\pi }{\beta }=\frac{\epsilon }{M^{2}}\ .
\label{a7}
\end{eqnarray}
The temperature of extreme RN black holes is zero as the regularization
parameter $\epsilon $ tends to zero.

It is worthwhile noting that when the general expression of temperature in
eq.(\ref{e16}) is applied to the extreme RN black hole, we obtain
\begin{eqnarray}
\left. \frac{2\pi }{\beta }\right\vert ^{{\rm Ext-RN}}=\left. 
\frac{\partial _{r}g_{\tau \tau}}{2\sqrt{g_{\tau \tau }g_{rr}}}\right
\vert_{r_{h} +\epsilon }^{{\rm Ext-RN}}=\frac{\epsilon }{M^{2}}\ ,
\label{a8}
\end{eqnarray}
which coincides the explicit calculation of eq.(\ref{a7}). We have applied
this prescription to define the temperature at the boundary position (brick
wall) of the scalar field. This value of $2\pi /\beta $ when combined with
the corresponding optical volume yields nontrivial results 
in the limit $\epsilon \rightarrow 0$, as we have seen in the text.

%%%%%%%%%%%%%%%%%%%%%%%%% references %%%%%%%%%%%%%%%%%%%

\end{document}